%% file: EHE_ICRC_2025.tex
\documentclass[a4paper,11pt]{article}

\usepackage{pos}
\usepackage{lipsum} %package to generate placeholder text in the following
\usepackage[separate-uncertainty]{siunitx}
\DeclareSIUnit\years{years}
\DeclareSIUnit\days{days}
\DeclareSIUnit\photoelectron{PE}

\title{A search for extremely-high-energy neutrinos with IceCube and implications for the ultra-high-energy cosmic-ray proton fraction}

\ShortTitle{Search for EHE$\nu$ with IceCube}

\author{The IceCube Collaboration \\{\normalsize \normalfont(a complete list of authors can be found at the end of the proceedings)}\\}

\emailAdd{mmeier@icecube.wisc.edu}
\emailAdd{brianclark@icecube.wisc.edu}

\abstract{

When ultra-high-energy cosmic rays (UHECRs) interact with ambient photon backgrounds, a flux of extremely-high-energy (EHE), so-called cosmogenic, neutrinos is produced. The observation of these neutrinos with IceCube can probe the nature of UHECRs. We present a search for EHE neutrinos using \SI{12.6}{\years} of IceCube data. The non-observation of neutrinos with energies $\gtrsim \SI{10}{\peta\electronvolt}$ constrains the all-flavor neutrino flux at \SI{1}{\exa\electronvolt} to be below $E^2 \Phi_{\nu_e + \nu_\mu + \nu_\tau} \simeq \SI{e-8}{\giga\electronvolt\per\square\centi\metre\per\second\per\steradian}$, the most stringent limit to date. This constrains the proton fraction in UHECRs of energy above \SI{30}{\exa\electronvolt} to be $\lesssim\SI{70}{\percent}$ if the evolution of the UHECR sources is similar to the star formation rate. Our analysis circumvents uncertainties associated with hadronic interaction models in studies of UHECR air showers, which also suggest a heavy composition at such energies.
\vspace{4mm}

{\bfseries Corresponding authors:}
Maximilian Meier$^{1*}$, 
Brian A. Clark$^{2}$\\
{$^{1}$ \itshape Chiba University}\\
{$^{2}$ \itshape University of Maryland, College Park}\\[4mm]
$^*$ Presenter
}

\input{ICRCdetails.tex}

\begin{document}

\maketitle

\section{Introduction}\label{sec:intro}

Extremely-high-energy (EHE, $E_{\nu} \gtrsim \SI{10}{\peta\electronvolt}$) neutrinos are unique messengers of the high redshift universe. Other Standard Model messenger particles do not arrive at Earth from high redshifts due to attenuation by their interactions with background photon fields. Neutrinos on the other hand carry no charge and only interact weakly, allowing them to reach Earth undeflected and unattenuated. EHE neutrinos are expected to be produced in interactions of ultra-high-energy cosmic rays (UHECRs): Either during their propagation through the universe, interacting with cosmic microwave background photons, called \textit{cosmogenic neutrinos}, or by cosmic-ray interactions in environments close to their astrophysical sources themselves, called \textit{astrophysical neutrinos}. The cosmogenic neutrino flux encodes unique information about the sources of UHECRs --- its shape and normalization can provide information about the chemical composition of UHECRs, the redshift evolution of their sources, and the maximum acceleration energy of the cosmic ray accelerators. In this work, we report on a search for EHE neutrinos based on \SI{12.6}{\years} of IceCube data~\cite{ehe_prl_2025}.

The IceCube Neutrino Observatory~\cite{icecube_instrumentation} is a cubic kilometer of deep ice at the geographic South Pole instrumented with \num{5160} Digital Optical Modules (DOMs), distributed on \num{86} strings, each housing a photomultiplier tube and readout electronics. Additionally, a surface array called IceTop~\cite{icetop}, deployed on top of the IceCube strings, measures cosmic-ray air showers. Neutrino interactions produce charged particles that give rise to Cherenkov light when they propagate through the ice. The Cherenkov light is recorded by the detector and can be used to reconstruct the arrival direction and energy of the neutrinos. EHE neutrinos are observed in IceCube as tracks --- $\mu$/$\tau$ leptons originating from charged-current (CC) interactions of $\nu_\mu$/$\nu_\tau$ depositing light while propagating through the ice ---  or cascades --- roughly spherical energy depositions from $\nu_\mathrm{e}$ CC or neutral-current interactions.

\section{Event Selection \label{sec:sel}}

The IceCube data is dominated by atmospheric muon bundles that trigger the detector at a rate of about \SI{3}{\kilo\hertz}, while the expected rate of cosmogenic neutrinos is constrained to be much smaller than 1 event per year. An event selection is applied to select high-energy neutrinos, focusing on neutrino energies greater than \SI{10}{\peta\electronvolt}, while rejecting atmospheric neutrino and muon backgrounds (for a detailed description, see~\cite{ehe_prl_2025}). Cosmogenic neutrino events have extremely high energy, depositing a large amount of light or equivalently charge. The atmospheric muon background is exclusively down-going, and thus a majority of the background can be removed with a charge threshold depending on the reconstructed arrival direction of the event. Another key difference between high-energy neutrino signals and atmospheric muon bundles is their energy loss profile. The energy loss profiles of single high-energy muons show large stochastic variations, while in high-multiplicity muon bundles these fluctuations partly average out. This information is included by reconstructing energy loss profiles along the reconstructed track directions and imposing more strict charge thresholds for less stochastic events. Finally, events are vetoed if they show activity in the IceTop surface array in temporal coincidence with the reconstructed track.

\begin{figure}
    \centering
    \includegraphics[width=0.65\linewidth]{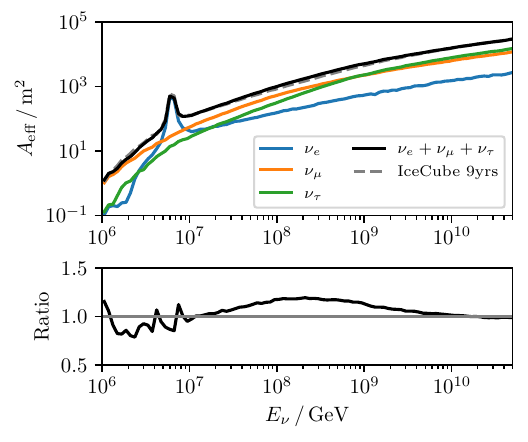}
    \caption{The $4\pi$-$\nu/\bar{\nu}$ averaged effective area of the event selection. The effective area of a previous search, IceCube \num{9}yrs from~\cite{ehe_9yr}, is shown as a dashed line for comparison.}
    \label{fig:effective_area}
\end{figure}

The effective area of the event selection is shown in Fig.~\ref{fig:effective_area} averaging the whole sky and the contributions from neutrinos and anti-neutrinos. This new event selection improves the effective area between \SI{100}{\peta\electronvolt} and \SI{1}{\exa\electronvolt} by about \SI{15}{\percent} due to the inclusion of stochasticity information. In addition to improvements in effective area, the analyzed livetime is increased by about \SI{50}{\percent} to a total of \SI{4605}{\days}.

After applying this event selection, the background expected in the analyzed livetime consists of \num{0.40 \pm 0.03} events from atmospheric sources and between $\sim$\num{9}~\cite{diffuse_numu} and $\sim$\num{0.5} ($\gamma = \num{2.39}, E_{\mathrm{cutoff}} = \SI{1.4}{\peta\electronvolt}$~\cite{globalfit_icrc}) astrophysical neutrinos depending on the assumed spectral behavior of astrophysical neutrinos. For the construction of upper limits described here, the parameters $\gamma = \num{2.39}$ and $E_{\mathrm{cutoff}} = \SI{1.4}{\peta\electronvolt}$ are assumed for astrophysical flux, described a power law with an exponential cutoff, because this leads to conservative results. Three data events, consistent with the expectation from astrophysical background, survive the event selection.

\section{Analysis and Results}\label{sec:results}

The sample is analyzed by splitting events into subsets made up of tracks and cascades and binning them based on their reconstructed energy and arrival direction. Then, a binned Poisson likelihood is used to fit the data:

\begin{equation}
    \mathcal{L}(\lambda_{\mathrm{GZK}}, \lambda_{\mathrm{astro}}) = \\ \prod_i \mathrm{Pois}(n_i | \lambda_{\mathrm{GZK}} \mu_{\mathrm{GZK}, i} + \lambda_{\mathrm{astro}} \mu_{\mathrm{astro}, i} + \mu_{\mathrm{bkg}, i}),
\end{equation}
where $\lambda_{\mathrm{GZK}}$ is a relative normalization to a signal flux of cosmogenic neutrinos and $\lambda_{\mathrm{astro}}$ is a nuisance parameter scaling the astrophysical neutrino flux. $n_i$ describes the observed number of events in bin $i$, and $\mu_i$ describes the expectation for a signal model, the astrophysical background and the sum of all atmospheric background respectively. A generic differential limit on the EHE neutrino flux beyond the astrophysical expectation is obtained by injecting a sliding $E^{-1}$ flux with a width of one decade. Additionally, specific signal models can also be tested using a likelihood-ratio test statistic. All hypothesis tests are based on ensembles of pseudoexperiments and confidence intervals are constructed based on the method described in~\cite{feldman_cousins}.

\begin{figure}
    \centering
    \includegraphics[width=0.65\linewidth]{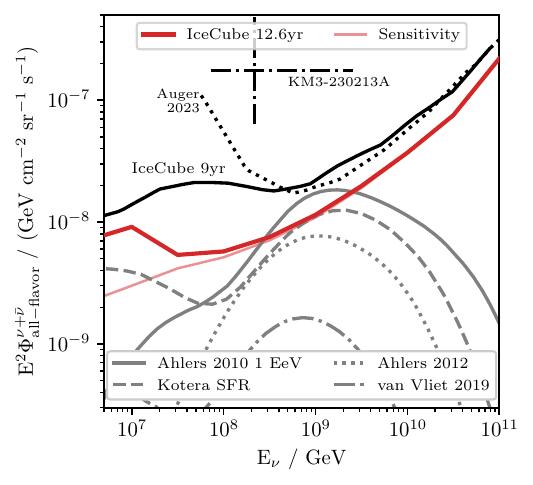}
    \caption{Differential limit at \num{90}\% CL on the all-flavor neutrino flux. The limit is compared to the IceCube result based on \SI{9}{\years} of data~\cite{ehe_9yr}, the limit by Auger~\cite{auger_limit} and to a few cosmogenic flux models~\cite{ahlers_gzk, Kotera:2010yn, ahlers_minimal, vanvliet}. The model labeled van Vliet 2019 assumes $\alpha = 2.5$, $E_{\mathrm{max}} = \SI{e20}{\electronvolt}$, $m = 3.4$ and a proton fraction of 10\%. The limit from Auger has been re-scaled to decade-wide bins for an easier comparison.} 
    \label{fig:diff_limit}
\end{figure}

The differential limit obtained in this work is shown in Fig.~\ref{fig:diff_limit} as a red line as well as the sensitivity, i.e. the limit obtained in case of a null observation. Below about \SI{100}{\peta\electronvolt} the limit is weakened with respect to the sensitivity due to the observed events.

Since the known flux of astrophysical neutrinos has been taken into account in the construction of the differential limit, the limit applies to any component of the UHE neutrino flux, whether it is cosmogenic neutrinos or a, so far unknown, component of UHE neutrinos produced directly in astrophysical sources.
% Maybe add some quantifying/interpreting statements here... and mention which astrophysical model is used

\subsection{Implications for the UHECR proton fraction}

Based on the observed flux of UHECRs, the non-observation of cosmogenic neutrinos imposes interesting constraints on the sources of UHECRs, in particular the UHECR proton fraction~\cite{ahlers_2009}. The connection between UHECRs and cosmogenic neutrinos is modeled using CRPropa simulation~\cite{crpropa3.2} and follows the method laid out in~\cite{vanvliet} with a detailed description given in~\cite{ehe_prl_2025}. The predicted cosmogenic neutrino flux depends mainly on the composition of UHECRs, the flux of UHECRs injected by their sources, and the source distribution as a function of redshift. The injected flux is modeled conservatively, where spectral parameters are chosen to minimize the expected neutrino flux. For the cosmological source evolution two different models are tested:

\begin{equation}
    \mathrm{SE}_1(z) = \begin{cases}
        (1 + z)^m, &z \leq z'\\
        (1 + z')^m, &z > z'
    \end{cases}
\end{equation}
with $z' = 1.5$ and $z_{\mathrm{max}} = 4$, and a more conservative model $\mathrm{SE}_2(z) = (1+z)^m$ with $z_{\mathrm{max}} = 2$, where $z$ denotes the redshift and $m$ the so-called source evolution parameter.

\begin{figure}
    \centering
    \includegraphics[width=0.75\linewidth]{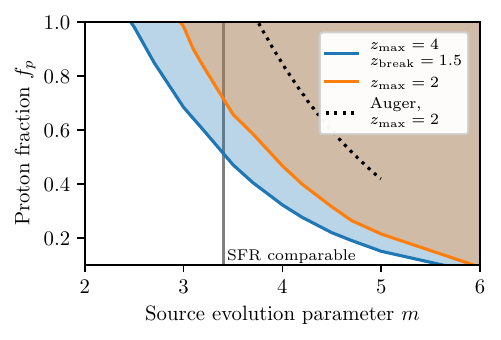}
    \caption{Constraints on the UHECR proton fraction as a function of the source evolution parameter $m$ at \num{90}\% CL. The excluded region is shown for the two described source evolution models $\mathrm{SE}_1(z)$ (blue), and $\mathrm{SE}_2(z)$ (orange), and compared to constraints based on the non-observation of neutrinos in Auger~\cite{auger_pfrac}.}
    \label{fig:pfrac}
\end{figure}

The resulting constraints on the proton fraction above $\sim\SI{30}{\exa\electronvolt}$ are shown in Fig.~\ref{fig:pfrac} as a function of the source evolution parameter. Since neutrinos can probe the distant universe, the proton fraction and the value of $m$ are degenerate. For instance, if the UHECR source evolution is comparable to the star formation rate (SFR) or stronger, the proton fraction is constrained to be below about \num{70}\%.

Direct air shower measurements already strongly constrain the composition of UHECRs to be heavy~\cite{auger_xmax, ta_isotropy}, and even favor the total absence of protons in the highest energy bin~\cite{auger_composition_icrc}. Still, some analyses find that an additional proton component that contributes up to $\sim$\num{10}\% to the total UHECR flux can improve the fit to the data compared to a one population model~\cite{ehlert, muzio}. The contribution can be reduced to the percent level by considering different hadronic interaction models. Regardless, in these scenarios the redshift evolution is not constrained when considering cosmic-ray observations alone. Neutrinos can effectively probe sources with large redshifts and a significant fraction of cosmogenic neutrinos originate from UHECR protons produced in sources with $z > 1$. This effectively allows for a measurement of the proton fraction as a function of $m$. For source classes with strong redshift evolution, such as high-luminosity AGN~\cite{agn_evolution} ($m = \num{7.1}$) the constraints derived in this work are competitive with the proton fractions allowed by cosmic-ray data.

\section{KM3-230213A}
Recently, the KM3NeT Collaboration published an EHE neutrino candidate with an estimated energy of $\sim\SI{220}{\peta\electronvolt}$~\cite{km3net_uhe}. This single event leads to an inferred diffuse neutrino flux in an energy range from \SI{72}{\peta\electronvolt} to \SI{2.6}{\exa\electronvolt} of $E^2\Phi_{\mathrm{all-flavor}} = \SI{1.74e-7}{\giga\electronvolt\per\square\centi\metre\per\second\per\steradian}$ assuming that the flux follows $E^{-2}$. The flux is also shown in Fig.~\ref{fig:diff_limit} and significantly exceeds the limits set in this work. This flux corresponds to an expectation of $\sim\num{70}$ events, which is inconsistent with a non-observation in the quoted energy range at the level of more than $\num{10}\sigma$.

This tension can be significantly reduced by considering the global picture of high-energy neutrino detectors. A joint fit combining the observation of KM3NeT with the non-observation of neutrinos in the same energy range by IceCube and Auger~\cite{auger_joint} reduces the tension based on the hypothesis of a diffuse flux to \num{2.6}$\sigma$~\cite{km3net_uhe, KM3NeT:2025ccp}. The joint fit in~\cite{km3net_uhe} uses the IceCube exposure from~\cite{ehe_9yr}. We repeat the joint fit including the IceCube exposure presented in this work. The probability of the joint fit flux of $E^2\Phi_{\mathrm{all-flavor}} = \SI{1.7e-9}{\giga\electronvolt\per\square\centi\metre\per\second\per\steradian}$ resulting in an observation with one event in KM3NeT, and no events in both Auger and IceCube is $\sim$\num{0.35}\%. The goodness-of-fit p-value is determined to be \num{0.4}\% (\num{2.9}$\sigma$) based on the saturated Poisson likelihood ratio test~\cite{baker_cousins}.

\begin{figure}
    \centering
    \includegraphics[width=0.65\linewidth]{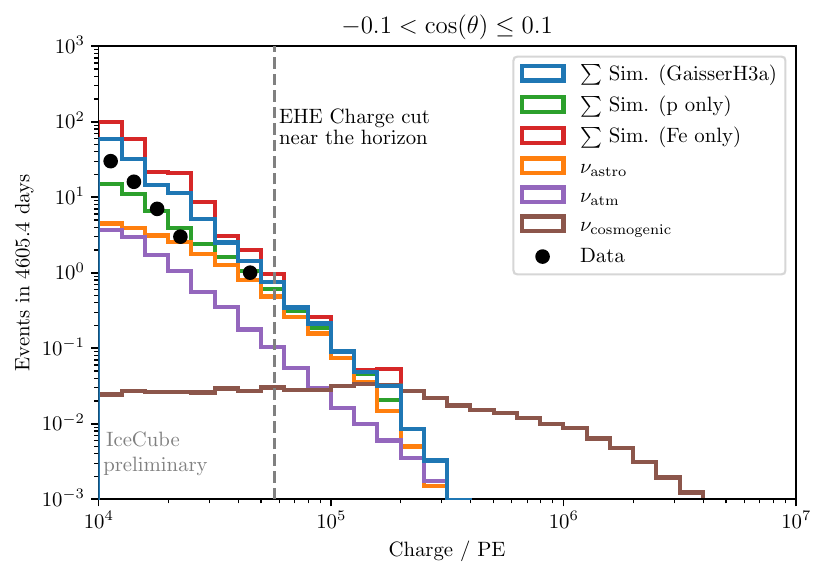}
    \caption{Charge distribution for events reconstructed close to the horizon with $-0.1 < \cos (\theta) \leq 0.1$. Except for the lower charge requirement of the histogram at $Q \geq \SI{e4}{\photoelectron}$ no selection criteria have been applied. The lines labeled $\sum \mathrm{Sim}$ combine the predictions for atmospheric muons, neutrinos and astrophysical neutrinos (assuming a single power law with cutoff: $\gamma = 2.52, E_{\mathrm{cutoff}} = \SI{1.4}{\peta\electronvolt}$~\cite{globalfit_icrc}). The distribution for cosmogenic neutrinos is plotted separately using the same model as the dash-dotted line in Fig.~\ref{fig:diff_limit}~\cite{vanvliet}.}
    \label{fig:horizontal_charge}
\end{figure}

This tension motivates to check whether IceCube is observing similarly spectacular near-horizontal events that could have been removed by the event selection summarized in Sec.~\ref{sec:sel}. Fig.~\ref{fig:horizontal_charge} shows a rather low-level distribution of the charge for events reconstructed to be close to horizontal with $\left| \cos(\theta) \right| \leq 0.1$. No other selection has been applied except for the lower charge requirement of $Q \geq \SI{e4}{\photoelectron}$. The simulation sum for different cosmic-ray primary composition models (GaisserH3a, proton only or iron only composition) is compared to the data represented by black points. No excess of events is observed above the combination of atmospheric backgrounds and astrophysical neutrinos. Additionally, the brown line shows the predicted charge distribution for a cosmogenic flux model. The distribution has a long tail to small charges, but an event similar to KM3-230213A where a muon with an energy of $\mathcal{O}(\SI{100}{\peta\electronvolt})$ passes through the detector with a favorable geometry can easily produce charges on the order of \SI{e6}{\photoelectron}.

\section{Conclusion}\label{sec:conclusion}

The non-observation of EHE neutrinos in \SI{12.6}{\years} of IceCube data places the strongest constraints to date on the flux of the highest energy neutrinos, reaching a flux at \SI{1}{\exa\electronvolt} of $E^2 \Phi \simeq \SI{e-8}{\giga\electronvolt\per\square\centi\metre\per\second\per\steradian}$. Additionally, this observation can constrain the UHECR composition, disfavoring a proton-only composition if the UHECR source evolution is comparable to or stronger than the star formation rate.

Planned projects utilizing the radio detection technique for neutrinos~\cite{gen2_tdr, grand} will reach neutrino fluxes about \num{1.5} orders of magnitude smaller than the current IceCube limit at \SI{1}{\exa\electronvolt}, probing cosmogenic neutrinos to significantly smaller proton fractions. The combination of a measurement of the cosmogenic neutrino flux with a proton fraction measurement from future cosmic ray detectors can be used to constrain the source evolution of potential pure-proton sources of UHECRs~\cite{muzio}.

% Bibtex references:
\bibliographystyle{ICRC}
\bibliography{references}

% Alternatively, you can include references by hand:
%\begin{thebibliography}{99}
%\bibitem{...}
%
%\end{thebibliography}

\clearpage

%The following list of authors, affiliations and funding agencies will be updated at the day of submission. The following template is a placeholder generated via https://authorlist.icecube.wisc.edu/icecube on May 17, 2025 and will be updated.
\input{authorlist_IceCube.tex}

\end{document}

%% file: ICRCdetails.tex
\FullConference{The 39th International Cosmic Ray Conference (ICRC2025)\\  14 -- 24 July, 2025\\ Geneva, Switzerland}

%% file: authorlist_IceCube.tex
\section*{Full Author List: IceCube Collaboration}

\scriptsize
\noindent
R. Abbasi$^{16}$,
M. Ackermann$^{63}$,
J. Adams$^{17}$,
S. K. Agarwalla$^{39,\: {\rm a}}$,
J. A. Aguilar$^{10}$,
M. Ahlers$^{21}$,
J.M. Alameddine$^{22}$,
S. Ali$^{35}$,
N. M. Amin$^{43}$,
K. Andeen$^{41}$
% ...
\\
\\
% ...
$^{10}$ Universit{\'e} Libre de Bruxelles, Science Faculty CP230, B-1050 Brussels, Belgium \\
% ...
$^{16}$ Department of Physics, Loyola University Chicago, Chicago, IL 60660, USA \\
$^{17}$ Dept. of Physics and Astronomy, University of Canterbury, Private Bag 4800, Christchurch, New Zealand \\
% ...
$^{21}$ Niels Bohr Institute, University of Copenhagen, DK-2100 Copenhagen, Denmark \\
$^{22}$ Dept. of Physics, TU Dortmund University, D-44221 Dortmund, Germany \\
% ...
$^{35}$ Dept. of Physics and Astronomy, University of Kansas, Lawrence, KS 66045, USA \\
% ...
$^{39}$ Dept. of Physics and Wisconsin IceCube Particle Astrophysics Center, University of Wisconsin{\textemdash}Madison, Madison, WI 53706, USA \\
% ...
$^{41}$ Department of Physics, Marquette University, Milwaukee, WI 53201, USA \\
% ...
$^{43}$ Bartol Research Institute and Dept. of Physics and Astronomy, University of Delaware, Newark, DE 19716, USA \\
% ...
$^{63}$ Deutsches Elektronen-Synchrotron DESY, Platanenallee 6, D-15738 Zeuthen, Germany \\
$^{\rm a}$ also at Institute of Physics, Sachivalaya Marg, Sainik School Post, Bhubaneswar 751005, India \\
% ...

\subsection*{Acknowledgments}

\noindent
The authors gratefully acknowledge the support from the following agencies and institutions:
USA {\textendash} U.S. National Science Foundation-Office of Polar Programs,
U.S. National Science Foundation-Physics Division,
U.S. National Science Foundation-EPSCoR,
U.S. National Science Foundation-Office of Advanced Cyberinfrastructure,
Wisconsin Alumni Research Foundation,
Center for High Throughput Computing (CHTC) at the University of Wisconsin{\textendash}Madison,
Open Science Grid (OSG),
Partnership to Advance Throughput Computing (PATh),
Advanced Cyberinfrastructure Coordination Ecosystem: Services {\&} Support (ACCESS),
Frontera and Ranch computing project at the Texas Advanced Computing Center,
U.S. Department of Energy-National Energy Research Scientific Computing Center,
Particle astrophysics research computing center at the University of Maryland,
Institute for Cyber-Enabled Research at Michigan State University,
Astroparticle physics computational facility at Marquette University,
NVIDIA Corporation,
and Google Cloud Platform;
Belgium {\textendash} Funds for Scientific Research (FRS-FNRS and FWO),
FWO Odysseus and Big Science programmes,
and Belgian Federal Science Policy Office (Belspo);
Germany {\textendash} Bundesministerium f{\"u}r Forschung, Technologie und Raumfahrt (BMFTR),
Deutsche Forschungsgemeinschaft (DFG),
Helmholtz Alliance for Astroparticle Physics (HAP),
Initiative and Networking Fund of the Helmholtz Association,
Deutsches Elektronen Synchrotron (DESY),
and High Performance Computing cluster of the RWTH Aachen;
Sweden {\textendash} Swedish Research Council,
Swedish Polar Research Secretariat,
Swedish National Infrastructure for Computing (SNIC),
and Knut and Alice Wallenberg Foundation;
European Union {\textendash} EGI Advanced Computing for research;
Australia {\textendash} Australian Research Council;
Canada {\textendash} Natural Sciences and Engineering Research Council of Canada,
Calcul Qu{\'e}bec, Compute Ontario, Canada Foundation for Innovation, WestGrid, and Digital Research Alliance of Canada;
Denmark {\textendash} Villum Fonden, Carlsberg Foundation, and European Commission;
New Zealand {\textendash} Marsden Fund;
Japan {\textendash} Japan Society for Promotion of Science (JSPS)
and Institute for Global Prominent Research (IGPR) of Chiba University;
Korea {\textendash} National Research Foundation of Korea (NRF);
Switzerland {\textendash} Swiss National Science Foundation (SNSF).